\pdfoutput=1
\documentclass[aps,prl,reprint,superscriptaddress,amsmath,amssymb,longbibliography]{revtex4-2}
\usepackage{graphicx}
\usepackage{bm}
\usepackage[colorlinks=true,linkcolor=blue,citecolor=blue,urlcolor=blue]{hyperref}
\newcommand{\dd}{\mathrm d}
\newcommand{\atanh}{\operatorname{artanh}}
\newcommand{\sech}{\operatorname{sech}}
\newcommand{\R}{\mathbb R}
\begin{document}

\title{Exactly solvable Brill waves: explicit gravitational-wave initial data\\ with trapped surfaces}
\author{Piotr Koc}
\email{piotr@koc.pl}
\affiliation{Independent researcher, Krak\'ow, Poland}
\date{November 21, 2026}

\begin{abstract}
Vacuum initial data for pure gravitational waves on $\mathbb{R}^3$ (Brill
waves) have so far been obtained only numerically. Here a one-parameter
family is solved exactly: in toroidal coordinates the constraint
separates, and a suitable seed function makes every mode equation
hypergeometric. The data are real analytic and asymptotically flat and exist
precisely for $k<k_c=0.8048\ldots$; their mass is an explicit series with a
pole at $k_c$. Near $k_c$ the data provably contain trapped surfaces and
an apparent horizon. A scaling symmetry yields further solvable families,
one of them elementary.
\end{abstract}
\maketitle

\emph{Introduction.}---In 1959 Dieter Brill, in his Princeton
dissertation and the paper based on it \cite{Brill}, studied axially
symmetric, time-symmetric vacuum initial data for the Einstein equations
and proved that their mass is positive. Although this is the simplest
nontrivial case of the vacuum constraints of general relativity, and the
standard setting for the old question whether gravitational waves alone
can form a black hole \cite{Wheeler}, no explicit solution has been
known. Over the last half-century an enormous effort has therefore
gone into solving the constraint numerically, for seed functions chosen
for convenience, in studies of the mass, of apparent horizons and of the
collapse of gravitational waves
\cite{Eppley,Holz,KKS,GarfinkleDuncan,Hilditch13,Hilditch17}, up to the
most recent work \cite{BGW25,BGIMW26,AranhaOliveira,MastersonHobill}.
Indirect arguments show that solutions exist for weak waves, that
existence is lost at a critical amplitude, and that trapped surfaces form
as this amplitude is approached \cite{CantorBrill,BeigOM}. The motivation
for seeking an explicit solution is that the vacuum data known in closed
form, namely the Schwarzschild, Brill--Lindquist and Misner data
\cite{BrillLindquist,Misner}, have their black holes built in through the
topology, while analytic Brill waves are known only in the linearized
theory \cite{LinearBrill}.

For time-symmetric data the vacuum constraints reduce to
${}^{3}\!R=0$, where ${}^{3}\!R$ is the scalar curvature of the initial
hypersurface. Brill's axially symmetric data \cite{Brill} are
\begin{equation}\label{metric}
 g=\Phi^4\bigl[e^{2q}(\dd\rho^2+\dd z^2)+\rho^2\dd\varphi^2\bigr],
 \qquad K_{ij}=0,
\end{equation}
where the seed function $q(\rho,z)$ is freely prescribed, subject to
the boundary conditions~\eqref{bc} below. For this metric the Hamiltonian
constraint takes the form of a Schr\"odinger equation at zero energy for
the conformal factor $\Phi>0$ and a Poisson equation in the $(\rho,z)$
plane, with the same function $V$ as potential, $-V$, and as source,
\begin{equation}\label{constraint}
 \Delta_3\Phi+V\Phi=0,\qquad
 V=\tfrac14\bigl(\partial_\rho^2+\partial_z^2\bigr)q ,
\end{equation}
where $\Phi=1+m/(2r)+O(r^{-2})$ at infinity, with $r^2=\rho^2+z^2$ and
$m$ the ADM mass; $\Delta_3$ is the flat Laplacian.
Regularity on the axis and asymptotic flatness require \cite{Brill}
\begin{equation}\label{bc}
 q=\partial_\rho q=0\ \ \text{at }\rho=0,\qquad
 q=O(r^{-2})\ \ (r\to\infty).
\end{equation}

The usual approach is to choose the seed function $q$ and then to
solve Eq.~\eqref{constraint} for $\Phi$, which has been possible only
numerically. This Letter takes the alternative approach: the potential
$V$ is chosen first, and Eqs.~\eqref{constraint} are then solved for
both $q$ and $\Phi$. The main problem is to choose $V$ so that this can
be done explicitly and the resulting $q$ is smooth and meets the
conditions~\eqref{bc}. It is solved by a separation of variables in
toroidal coordinates, described first, with a potential for which every
separated equation is hypergeometric. The result is a one-parameter
family of real-analytic data with trivial topology and a single end: the
first explicit vacuum initial data for the Einstein equations that are
smooth on all of $\R^3$, asymptotically flat and of positive mass. The
family has an exact critical parameter, an exact mass formula and,
provably, trapped surfaces, and it provides an exact benchmark for
constraint solvers, mass computations and horizon finders, including an
elementary member, Eq.~\eqref{simplest}.
Proofs are given in a companion paper \cite{companion}.

\begin{figure*}
\includegraphics[width=\textwidth]{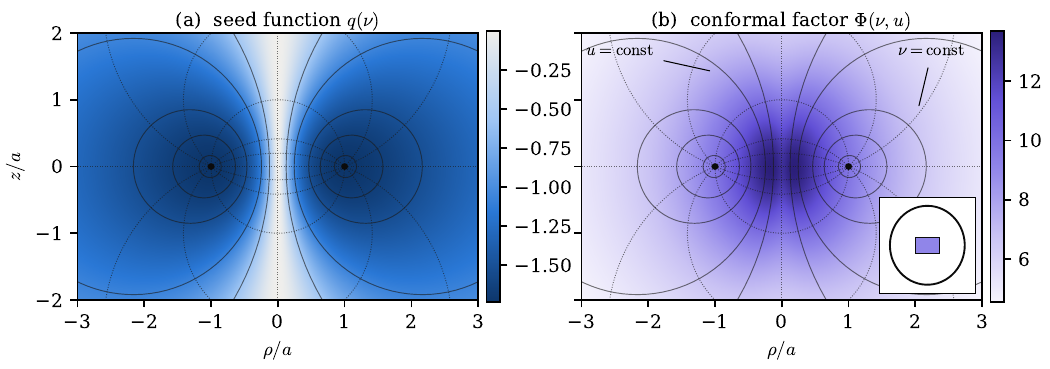}
\caption{\label{fig1}The member $k=0.78$ of the family in a meridional
plane, with $\rho$ continued to negative values across the axis.
(a)~The seed function $q$ depends on $\nu$ only and is constant on the
tori $\nu=\mathrm{const}$ (solid circles). (b)~The conformal factor
$\Phi$, Eq.~\eqref{Phi}, depends on both $\nu$ and $u$; dotted circles
are surfaces $u=\mathrm{const}$. Dots mark the circle $\rho=a$, $z=0$,
where $\Phi$ takes the value \eqref{circle}. Here $m=21.7\,a$. Inset:
the outer marginally trapped sphere, at $r\approx9.4$--$9.9\,a$, around
the frame of the panels (rectangle).}
\end{figure*}

\emph{Separation of variables.}---Toroidal coordinates
$(\nu,u,\varphi)$ adapted to a circle of radius $a$ are defined by
\begin{equation}\label{coords}
 \rho=\frac{a\sinh\nu}{d},\quad z=\frac{a\sin u}{d},\quad
 d(\nu,u)=\cosh\nu-\cos u,
\end{equation}
with $0\le\nu<\infty$ and $-\pi<u\le\pi$; the function $d(\nu,u)$ is
written $d$ for short. The surfaces
$\nu=\mathrm{const}$ are tori around the circle $\rho=a$, $z=0$, which
is reached as $\nu\to\infty$; the axis is $\nu=0$, and spatial infinity
is the single point $(\nu,u)=(0,0)$ (Fig.~\ref{fig1}). The map is
conformal, and in these coordinates the two equations of
\eqref{constraint}, divided by $d^2/a^2$, read (End Matter)
\begin{gather}
 \begin{split}
 \Phi_{\nu\nu}+\coth\nu\,\Phi_\nu+\Phi_{uu}
 -\frac{\sinh\nu\,\Phi_\nu+\sin u\,\Phi_u}{d}&\\
 +\frac{a^2}{d^2}V\,\Phi&=0,
 \end{split}\label{Phitor}\\
 \frac{a^2}{d^2}\,V=\tfrac14\bigl(q_{\nu\nu}+q_{uu}\bigr).\label{qtor}
\end{gather}
The boundary conditions \eqref{bc} become (End Matter)
\begin{equation}\label{bctor}
\begin{split}
 &q=q_\nu=\Phi_\nu=0\quad\text{at }\nu=0,\\
 &\Phi=1+\frac{m\sqrt d}{2\sqrt2\,a}+O(d),\quad q=O(d)\quad\text{at }(\nu,u)=(0,0),
\end{split}
\end{equation}
and $q$, $\Phi$ must be smooth at the circle $\nu\to\infty$, a
condition with no counterpart in Eq.~\eqref{bc}.
The separation rests on three observations.

(i) \emph{A seed that depends on $\nu$ alone.} We assume that $q$
depends on $\nu$ only, $q=q(\nu)$. The combination $(a^2/d^2)V$ appears
in both equations; call it the reduced potential, $w\equiv(a^2/d^2)V$.
Then Eq.~\eqref{qtor} becomes
\begin{equation}\label{V}
 w=\tfrac14q''(\nu),
\end{equation}
a function of $\nu$ alone. This relation will be read backwards: $w$ is
prescribed, and $q=4\int_0^\nu\!\dd\nu'\!\int_0^{\nu'}\!w\,\dd\nu''$
follows, with the boundary conditions on $q$ imposed on the result.

(ii) \emph{Removing the factor $d$.} The coefficients of
Eq.~\eqref{Phitor} now depend on $u$ only through the single term
$-(\sinh\nu\,\Phi_\nu+\sin u\,\Phi_u)/d$. This term is removed by the
substitution $\Phi=\sqrt d\,F(\nu,u)$, because of the identity
\begin{equation}\label{identity}
 \Delta_3\bigl(\sqrt d\,F\bigr)=\frac{d^{5/2}}{a^2}
 \Bigl[F_{\nu\nu}+\coth\nu\,F_\nu+F_{uu}+\tfrac14F\Bigr],
\end{equation}
which also underlies the toroidal harmonics of potential theory (End
Matter). With Eq.~\eqref{V}, the constraint \eqref{Phitor} is
therefore equivalent to
\begin{equation}\label{Feq}
 F_{\nu\nu}+\coth\nu\,F_\nu+F_{uu}+\bigl[\tfrac14+w(\nu)\bigr]F=0 ,
\end{equation}
an equation whose coefficients do not depend on $u$.

(iii) \emph{Fourier modes.} For data symmetric about the plane $z=0$,
$F$ is periodic and even in $u$. Writing
\begin{equation}\label{Fourier}
 F(\nu,u)=\sum_{n=0}^\infty f_n(\nu)\cos(nu)
\end{equation}
turns Eq.~\eqref{Feq} into ordinary differential equations,
\begin{equation}\label{modes}
 f_n''+\coth\nu\,f_n'+\bigl[\tfrac14-n^2+w(\nu)\bigr]f_n=0 .
\end{equation}
For $w=0$ these are Legendre equations, and $\Phi=1$ is the classical
expansion of $1/\sqrt d$ in the functions $Q_{n-1/2}(\cosh\nu)$
\cite{Cohl}.

This reduction is due to Beig and Husa \cite{BeigHusa}, who identified
the underlying conformal symmetry and used it in numerical constructions
\cite{Husa98,Husa96}. Geometrically, the seed metric is conformal, with factor
$d^2/(a^2\cosh^2\nu)$, to a metric on $S^3$ whose coefficients do not
depend on $u$; the constraint is conformally covariant, and $\sqrt d$
is, up to a function of $\nu$, the corresponding conformal weight. What
is new here is a potential $w$ for which all of Eqs.~\eqref{modes} are
solved in closed form.

\emph{A solvable family.}---Following the strategy stated in the
introduction, $w$ is now chosen so that Eqs.~\eqref{modes} are solvable
in closed form; $q$ is obtained afterwards from Eq.~\eqref{V}, and the
boundary conditions on it fix the one constant that solvability leaves
free. This strategy was introduced in \cite{Koc21} within a thin-torus
approximation, whose results are superseded by the exact family below.
With $f_n=g_n/\sqrt{\sinh\nu}$,
Eq.~\eqref{modes} becomes a one-dimensional Schr\"odinger equation at
zero energy,
\begin{equation}\label{schrodinger}
 -g_n''+\Bigl[n^2-\frac1{4\sinh^2\nu}-w(\nu)\Bigr]g_n=0 .
\end{equation}
Natanzon \cite{Natanzon} described the potentials for which such
equations are reduced to the hypergeometric equation by a change of
variable. The construction leaves no freedom beyond a few constants:
demanding that the transformed equation be exactly the hypergeometric
equation, for every $n$ with one and the same new variable, fixes both
the variable and the potential, up to the scaling of $\nu$ used below.
With the axis and the circle as the
two finite singular points, approached as regularity there requires,
the variable is $\sigma\in[0,1)$, with a real constant $k<1$, defined by
\begin{equation}\label{sigma}
 \frac{\dd\sigma}{\dd\nu}=\frac{1-\sigma^2}{R},\qquad
 R=\sqrt{1-k+k\sigma^2},\qquad k<1,
\end{equation}
where $\sigma=0$ is the axis and $\sigma\to1$ the circle.
Equation~\eqref{sigma} integrates to
\begin{equation}\label{nusigma}
 \nu=\atanh\frac{\sigma}{R}-\sqrt k\,\atanh\frac{\sqrt k\,\sigma}{R};
\end{equation}
the inverse $\sigma(\nu)$ is elementary only for $k=0$, where
$\sigma=\tanh\nu$, so all formulas below are written parametrically in
$\sigma$. With this variable, the same requirement---that the
transformed equation be exactly the hypergeometric equation---forces
the potential to be
\begin{equation}\label{w}
\begin{split}
 w={}&\frac{S(1-\sigma^2)}{R^2}
 +\frac{(1-\sigma^2)\bigl[(1-k)^2(1-\sigma^2)+4k\sigma^4\bigr]}
       {4\sigma^2R^6}\\
 &-\frac1{4\sinh^2\nu},
\end{split}
\end{equation}
with a single real constant $S$ as the only freedom left. The constant
$S$ is fixed by regularity at the circle. For $q=q(\nu)$ the axis
conditions in \eqref{bctor} are $q(0)=q'(0)=0$, so
$q'(\nu)=4\int_0^\nu w\,\dd\nu'$. The decay condition then holds as
well, because $q\simeq2w(0)\nu^2=O(d)$ at $(0,0)$. Near the circle
$\nu\simeq\ln(2a/\delta)$, with $\delta$ the distance from it, so a
non-zero limit $q'(\infty)$ would make $q$ logarithmically singular;
smoothness of $q$ at the circle thus implies $q'(\infty)=0$, that is,
$\int_0^\infty w\,\dd\nu=0$. This is
the same as $\int V\,\dd^3x=0$, the zero-average property of the
potential of regular Brill waves \cite{Wheeler}. In Eq.~\eqref{w} the
constant $S$ enters linearly, and with $\dd\nu=R\,\dd\sigma/(1-\sigma^2)$
the integral evaluates to $S\,\atanh\sqrt k/\sqrt k-(3+k)/[12(1-k)]$;
its vanishing gives
\begin{equation}\label{S}
 S(k)=\frac{(3+k)\sqrt k}{12(1-k)\atanh\sqrt k},
\end{equation}
which continues to $k<0$, where $\atanh\sqrt k/\sqrt k$ becomes
$\arctan\sqrt{|k|}/\sqrt{|k|}$. The single
parameter $k$ measures the strength of the wave; $k=0$ is flat space,
with $S=\tfrac14$ and $w=0$.

\emph{The seed function.}---Two integrations of Eq.~\eqref{V} give
$q$ in closed form: with $c=(1-\sqrt k)/(1+\sqrt k)$,
$t=(R-\sigma)/(R+\sigma)$, the dilogarithm $\mathrm{Li}_2$ and
\begin{equation}\label{dilog}
 \mathcal L=\mathrm{Li}_2(-ct)-\mathrm{Li}_2(-c)
 -\mathrm{Li}_2(-t/c)+\mathrm{Li}_2(-1/c),
\end{equation}
the seed function with $q(0)=q'(0)=0$ is
\begin{equation}\label{qclosed}
\begin{split}
 q={}&\ln\frac{\sinh\nu}{\sqrt{1-k}\,\sigma}
   -\frac56\ln\frac{R^2}{1-k}
   +\frac{3-k}{3(1-k)}\ln(1-\sigma^2)\\
   &+\frac{S}{\sqrt k}\Bigl[4\,\atanh\frac{\sigma}{R}\,\atanh\sqrt k
     +\mathcal L\Bigr]\\
   &-2S\,\atanh^2\frac{\sqrt k\,\sigma}{R} .
\end{split}
\end{equation}
For $k<0$ one takes $\sqrt k=i\sqrt{|k|}$ and principal branches; the
result is real. For $k=-3$, where $S=0$, the dilogarithms drop out:
\begin{equation}\label{qm3}
 q=\ln\frac{\sinh\nu}{2\sigma}-\frac56\ln\frac{4-3\sigma^2}{4}
   +\frac12\ln(1-\sigma^2).
\end{equation}
For every $k<1$, $q$ and $V$ are real analytic on $\R^3$, and $q$
decays as $r^{-2}$.

\emph{The conformal factor.}---We now return to the mode
equations~\eqref{modes}. With the potential
\eqref{w}, the change of variable $\nu\to x=1-\sigma^2$ defined by
Eq.~\eqref{sigma}, and the substitution
$f_n=\sqrt{\sigma R/\sinh\nu}\,(1-\sigma^2)^{n/2}\,y_n(x)$, every one
of them becomes
\begin{equation}\label{hypergeom}
 x(1-x)\,y_n''+\bigl[(n+1)-(n+2)x\bigr]y_n'-\alpha_n\beta_n\,y_n=0,
\end{equation}
the Gauss hypergeometric equation with parameters $\alpha_n,\beta_n$
and $n+1$, where
$\alpha_n+\beta_n=n+1$ and $\alpha_n\beta_n=\tfrac14[(n+1)^2-S-kn^2]$.
The solution that is regular at the circle, $x=0$, is
\begin{equation}\label{fn}
\begin{split}
 f_n={}&\frac{\epsilon_n}{\sqrt2\,\pi}\,
   \frac{\Gamma(\alpha_n)\Gamma(\beta_n)}{n!}\,
   \sqrt{\frac{\sigma R}{\sinh\nu}}\,(1-\sigma^2)^{n/2}\\
 &\times{}_2F_1\bigl(\alpha_n,\beta_n;n+1;1-\sigma^2\bigr),
\end{split}
\end{equation}
with $\alpha_n,\beta_n=\tfrac12(1+n\mp s_n)$, $s_n=\sqrt{S+kn^2}$,
$\epsilon_0=1$ and $\epsilon_n=2$ for $n\ge1$. The normalization factor
$\epsilon_n\Gamma(\alpha_n)\Gamma(\beta_n)/(\sqrt2\,\pi\,n!)$ follows
from the condition $\Phi=\sqrt d\,F\to1$ at infinity. Undoing the
substitution of step (ii) and using the expansion
\eqref{Fourier}, the conformal factor is
\begin{equation}\label{Phi}
 \Phi=\sqrt d\,\sum_{n=0}^\infty f_n(\nu)\cos(nu).
\end{equation}
Equations \eqref{w}--\eqref{Phi} are the solution: no thin-torus or
weak-field approximation is involved. On the circle only the mode $n=0$
contributes, and
\begin{equation}\label{circle}
 \Phi\big|_{\rho=a,\,z=0}=\frac1{\sqrt2\,\cos\bigl(\pi\sqrt S/2\bigr)} .
\end{equation}

\emph{A scaling symmetry.}---Under
$\nu\to\ell\nu$ with an integer $\ell$, the only term of
Eq.~\eqref{schrodinger} that does not scale is $1/(4\sinh^2\nu)$, and
its double pole at the axis is scale invariant. Consequently
\begin{equation}\label{wl}
 w_\ell(\nu)=\ell^2\Bigl[w(\ell\nu)+\frac1{4\sinh^2(\ell\nu)}\Bigr]
 -\frac1{4\sinh^2\nu}
\end{equation}
is smooth at the axis and analytic in $e^{-2\nu}$ at the circle whenever
$w$ is, and its modes are Eq.~\eqref{fn} with $n/\ell$ in place of $n$,
$\sigma(\ell\nu)$ in place of $\sigma$, and $\ell\sinh\nu$ in place of
$\sinh\nu$. Applied to Eq.~\eqref{w} with $S$ free, this gives for every
$\ell$ a family in $k$, with
\begin{equation}\label{Sl}
\begin{split}
 S&=\frac{\sqrt k}{\atanh\sqrt k}\Bigl[\frac{3+k}{12(1-k)}+\frac{\ell-1}{4\ell}\Bigr],\\
 q&=\tilde q(\ell\nu)+\ln\frac{\ell\sinh\nu}{\sinh(\ell\nu)},
\end{split}
\end{equation}
where the first line is the circle condition and $\tilde q$ is the closed
form \eqref{qclosed} at that $S$.

\emph{The simplest explicit wave.}---The member $\ell=2$, $k=0$ is the
simplest explicit Brill wave known to the author:
\begin{equation}\label{simplest}
\begin{split}
 q&=\tfrac12\ln\bigl(1+\tanh^2\nu\bigr),\\
 V&=\frac{d^2}{a^2}\Bigl[\tfrac12\sech^2(2\nu)-\tfrac14\sech^2\nu\Bigr],\\
 \Phi&=\sqrt d\sum_{n\ge0}
 \frac{\epsilon_n\Gamma(\tilde\alpha_n)\Gamma(\tilde\beta_n)}{\sqrt2\,\pi\,\Gamma(\tfrac n2+1)}
 \sqrt{\frac{\tanh2\nu}{2\sinh\nu}}\,(\sech2\nu)^{\frac n2}\\
 &\qquad\times{}_2F_1\bigl(\tilde\alpha_n,\tilde\beta_n;\tfrac n2+1;\sech^22\nu\bigr)\cos(nu),
\end{split}
\end{equation}
with $\tilde\alpha_n,\tilde\beta_n=(4+2n\mp\sqrt6)/8$ and $S=\tfrac38$.
Through $\tanh\nu=2a\rho/(r^2+a^2)$, $q$ is rational-logarithmic and
$V$ rational in the cylindrical coordinates; the mass is $m=0.0394\,a$,
and $\Phi=0.874$ on the circle \cite{companion}. The wave contains no
trapped surfaces: every sphere $r=\mathrm{const}$ has positive mean
curvature, so a closed surface with $H\le0$ would touch its smallest
enclosing sphere from inside, which the maximum principle forbids
\cite{companion}.

\emph{Existence and the critical parameter.}---A series solution is of
use only if it represents a global, positive solution. This is settled
by the function
$h_k=\sqrt d\,\sqrt{\sigma R/\sinh\nu}$, which is smooth and positive
on $\R^3$, decays as $1/r$, and satisfies
\begin{equation}\label{h}
 (\Delta_3+V)h_k=-(1-S)\,C_kh_k,\qquad
 C_k=\frac{d^2(1-\sigma^2)}{a^2R^2}>0 .
\end{equation}
It yields the ground-state identity
$\int(|\nabla\eta|^2-V\eta^2)=\int h_k^2|\nabla(\eta/h_k)|^2
+(1-S)\int C_k\eta^2$ for test functions $\eta$, so the constraint
operator is coercive exactly when $S<1$. As $V=S\,C_k+\dots$, the
constant $S$ measures the depth of the potential well $-V$ of
Eq.~\eqref{constraint}, and $S=1$ is exactly the threshold at which
the well begins to bind.

\emph{Theorem 1 \cite{companion}.} For $k<1$, a positive solution of
Eq.~\eqref{constraint} with $\Phi\to1$ exists if and only if $S(k)<1$,
that is, for $k<k_c$, where $k_c$ is the unique root of
\begin{equation}\label{kc}
 12(1-k_c)\frac{\atanh\sqrt{k_c}}{\sqrt{k_c}}=3+k_c,\qquad
 k_c=0.8048016772\ldots
\end{equation}
The solution is unique, real analytic and given by Eq.~\eqref{Phi}; the
metric \eqref{metric} is complete and asymptotically flat.

At $k=k_c$ the function $h_{k_c}$ is a zero mode of the constraint
operator, a bound state at zero energy, and the value \eqref{circle}
diverges. The limit of existence,
which in numerical work is bracketed by trial \cite{MastersonHobill}, is
here the root of an elementary equation.

\begin{figure}
\includegraphics[width=\columnwidth]{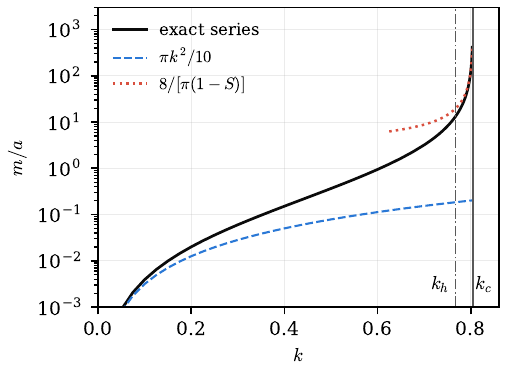}
\caption{\label{fig2}ADM mass of the family for $k>0$. Solid: the exact
series \eqref{mass}. Dashed: the weak-field law \eqref{weak}. Dotted:
the pole \eqref{pole}. Marginally trapped spheres are found numerically
for $k>k_h$; solutions exist for $k<k_c$.}
\end{figure}

\emph{Mass.}---The ADM mass is read off from the behavior of the modes
at the axis. With the digamma function $\psi=\Gamma'/\Gamma$,
Eq.~\eqref{fn} gives
\begin{equation}\label{mass}
\begin{split}
 \frac ma=\frac4\pi\sum_{n=0}^\infty\epsilon_n\Bigl[&\psi\bigl(n+\tfrac12\bigr)
  -\frac{\psi(\alpha_n)+\psi(\beta_n)}2\\
  &+\frac12\ln\frac{1-k}4\Bigr].
\end{split}
\end{equation}
The terms are $O(n^{-2})$ and the series converges absolutely.
For weak waves,
\begin{equation}\label{weak}
 \frac ma=\frac{\pi}{10}\,k^2+O(k^3),
\end{equation}
and at the critical point the mass has a simple pole,
\begin{equation}\label{pole}
 \frac ma=\frac{8}{\pi\,[1-S(k)]}+O(1)\qquad(k\uparrow k_c),
\end{equation}
with residue $8/[\pi S'(k_c)]=0.629$ in the variable $k$
(Fig.~\ref{fig2}). The mass blow-up at the loss of positivity of the
constraint operator is a general phenomenon \cite{BeigOM}; here it is
obtained in closed form, where numerical studies fit the dependence of
the mass on the amplitude \cite{AranhaOliveira}. Representative values
are collected in Table~\ref{tab1}.

\begin{table}[b]
\caption{\label{tab1}Members of the family: the constant \eqref{S}, the
mass \eqref{mass}, the conformal factor on the circle \eqref{circle},
and, where a marginally trapped sphere was found, the ratio
$\sqrt{\mathcal A/16\pi}/m$ for the outer one. The row $k=0.7670$ is
the fold \eqref{fold}.}
\begin{ruledtabular}
\begin{tabular}{lcccc}
$k$&$S$&$m/a$&$\Phi_{\rm circle}$&$\sqrt{\mathcal A/16\pi}/m$\\
\hline
$-3$&$0$&$0.2311$&$0.7071$&---\\
$0.5$&$0.4680$&$0.3660$&$1.485$&---\\
$0.7$&$0.7107$&$3.202$&$2.897$&---\\
$0.7670$&$0.8693$&$13.13$&$6.669$&$0.9713$\\
$0.78$&$0.9097$&$21.66$&$9.752$&$0.9951$\\
$0.80$&$0.9810$&$127.0$&$47.11$&$0.99999$\\
\end{tabular}
\end{ruledtabular}
\end{table}

\emph{Trapped surfaces.}---For time-symmetric data a closed surface is
marginally trapped when it is minimal, and outer trapped when its mean
curvature with respect to the outer normal is negative. The existence
of trapped surfaces near the critical amplitude was established by Beig
and \'O~Murchadha \cite{BeigOM}; the present family is the first in
which this mechanism can be followed explicitly, with an exact critical
parameter and trapped surfaces exhibited in closed form. This is
possible because only the mode $n=0$ is singular at $k_c$.

\emph{Theorem 2 \cite{companion}.} Let $A_k$ be the value
\eqref{circle}. As $k\uparrow k_c$,
\begin{equation}\label{profile}
 \frac{\Phi_k}{A_k}\longrightarrow h_{k_c}
 \quad\text{in }C^2_{\rm loc}(\R^3),
\end{equation}
and for $k$ sufficiently close to $k_c$ the data contain outer trapped
tori and outer trapped spheres, and therefore an apparent horizon.

The limit \eqref{profile} has a simple geometric meaning. The rescaled
metrics $A_k^{-4}g$ converge to
$g_\infty=h_{k_c}^4[e^{2q}(\dd\rho^2+\dd z^2)+\rho^2\dd\varphi^2]$,
which is scalar-flat, is invariant under the inversion $r\to a^2/r$, and
extends smoothly over the point at infinity to a metric on $S^3$: the
wave nearly pinches off a closed universe, joined to the asymptotically
flat region by a neck near $r=m/2$. In $g_\infty$ the tori
$\nu=\nu_0<0.371$ and the coordinate spheres $r\ge5.45\,a$ have negative
mean curvature. Since the mean curvature involves only
$\nabla\ln\Phi$, the constant $A_k$ drops out, and these surfaces are
outer trapped in the actual data for $k$ close to $k_c$. Barrier
arguments \cite{AnderssonMetzger,Eichmair,HuiskenIlmanen} then give an
outermost minimal surface, an apparent horizon, enclosing them. Trapped
tori of this kind were studied in \cite{Husa96,KarkowskiTori}.

Theorem~2 gives no estimate of how close to $k_c$ one must be. A
numerical search on the exact data, restricted to equatorially
symmetric star-shaped surfaces, shows that marginally trapped spheres
exist well before the critical point: they appear in a fold at
\begin{equation}\label{fold}
 k_h\simeq0.7670,\qquad m(k_h)\simeq13.13\,a .
\end{equation}
Of the two branches, the outer one approaches the Schwarzschild value
$r=m/2$, and its area $\mathcal A$ satisfies the Penrose inequality
\cite{HuiskenIlmanen,Bray}, with a deficit of $2.9\%$ at the fold that shrinks rapidly
towards $k_c$ (Table~\ref{tab1}). The inner
branch tends to the sphere $r=a$, which is minimal in $g_\infty$. This
search does not exclude other marginally trapped surfaces, and no claim
is made about the absence of horizons for $0<k<k_h$. Below $k_h$ the
curvature is already large: at the center the Kretschmann scalar reaches
$\mathcal Km^4=786$, against $3/4$ on a Schwarzschild horizon
\cite{companion}.

The branch $k<0$ has no critical point, since \mbox{$S(k)<1$} there, and
its mass grows only slowly, to $1.25\,a$ at \mbox{$k=-100$}. For $-8\le k<0$ every
sphere $r=\mathrm{const}$ has positive mean curvature, which excludes
trapped surfaces as for the simplest wave; for lower $k$, examined down
to $k=-30$, no sphere $r=\mathrm{const}$ or torus $\nu=\mathrm{const}$
is trapped.

\emph{Discussion.}---The quantities that are computed numerically for
Brill waves, namely the conformal factor, the mass, the limit of
existence and the formation of trapped surfaces, are obtained here in
exact form, for a particular family of seed functions. The price is
that the seed is dictated by solvability: it has the toroidal conformal
symmetry of Ref.~\cite{BeigHusa} and is not one of the profiles
commonly used in simulations. Within this limitation the family
provides exact benchmarks and exact initial data for evolution codes. For
$k_h<k<k_c$ the data are pure vacuum, topologically trivial and
contain marginally trapped surfaces; as argued in Ref.~\cite{BeigOM}, the
spacetimes evolving from such data must develop singularities. How the
exact thresholds $k_h$ and $k_c$ of the initial data relate to the
threshold of black-hole formation in the evolution
\cite{Hilditch13,Hilditch17}, and whether solvable potentials exist
beyond the hypergeometric class of Eqs.~\eqref{w} and \eqref{wl}, are
natural questions for future work.

\begin{acknowledgments}
Large language model assistants, including Claude (Anthropic), were
used in the exploration that led to the solvable family, in symbolic
and numerical checks, and in drafting and revising the text. The author
has reviewed the results and the text and takes full responsibility for
the content of this article. The Python code that reproduces all
numbers and figures is available with the companion paper
\cite{companion}.
\end{acknowledgments}

\section*{End Matter}

\emph{Appendix A: The separation identity.}---Since $\rho+iz
=a\coth[(\nu-iu)/2]$, the map $(\nu,u)\mapsto(\rho,z)$ is conformal and
the flat metric is
$(a^2/d^2)(\dd\nu^2+\dd u^2+\sinh^2\nu\,\dd\varphi^2)$, with scale
factors $h_\nu=h_u=a/d$ and $h_\varphi=a\sinh\nu/d$. The general formula
for the Laplacian in orthogonal coordinates gives, for functions
independent of $\varphi$,
\begin{equation}\label{Lap}
\begin{split}
 \Delta_3\Phi&=\frac{d^3}{a^2\sinh\nu}\Bigl[
   \partial_\nu\Bigl(\frac{\sinh\nu}{d}\Phi_\nu\Bigr)
  +\partial_u\Bigl(\frac{\sinh\nu}{d}\Phi_u\Bigr)\Bigr]\\
 &=\frac{d^2}{a^2}\Bigl[\Phi_{\nu\nu}+\coth\nu\,\Phi_\nu+\Phi_{uu}\\
 &\qquad\qquad
  -\frac{\sinh\nu\,\Phi_\nu+\sin u\,\Phi_u}{d}\Bigr],
\end{split}
\end{equation}
the second form by $\partial_\nu(1/d)=-\sinh\nu/d^2$ and
$\partial_u(1/d)=-\sin u/d^2$; dividing $\Delta_3\Phi+V\Phi=0$ by
$d^2/a^2$ gives Eq.~\eqref{Phitor}. Conformal covariance of the
two-dimensional Laplacian gives
$\partial_\rho^2+\partial_z^2=(d^2/a^2)(\partial_\nu^2+\partial_u^2)$,
hence Eq.~\eqref{qtor}. On the axis $\rho\simeq a\nu/(1-\cos u)$, so
$\partial_\rho\propto\partial_\nu$ there; at $(\nu,u)=(0,0)$ one has
$r^2/a^2=(\cosh\nu+\cos u)/d\simeq2/d$, so $m/(2r)\simeq m\sqrt d/(2\sqrt2\,a)$
and $O(r^{-2})=O(d)$; this gives Eq.~\eqref{bctor}. For the identity
\eqref{identity}, insert $\Phi=\sqrt d\,F$ in the first form of
Eq.~\eqref{Lap} and use $\partial_\nu d=\sinh\nu$,
$\partial_ud=\sin u$. The first-order terms that contain $1/d$ cancel,
and the terms without derivatives of $F$ combine to
$\sinh\nu\,d^{-5/2}F\,[d\cosh\nu+\tfrac12d\cos u
-\tfrac34(\sinh^2\nu+\sin^2u)]$. Since
$\sinh^2\nu+\sin^2u=d\,(\cosh\nu+\cos u)$, the bracket equals $d^2/4$,
which gives Eq.~\eqref{identity}.

\emph{Appendix B: Sketch of the proof of Theorem~2.}---For $n=0$ one
has $\beta_0=1-\alpha_0$, and Eq.~\eqref{fn} gives exactly
$\sqrt d\,f_0=A_kh_k\,{}_2F_1(\alpha_0,1-\alpha_0;1;1-\sigma^2)$, where
$A_k=[\sqrt2\sin(\pi\alpha_0)]^{-1}$ is the value \eqref{circle} and
$\alpha_0\to0$ as $k\uparrow k_c$. The modes with $n\ge1$ are positive
and are bounded, uniformly in $k$, by Legendre functions of the second
kind through a Wronskian comparison. Hence $\Phi_k-A_kh_k$ stays bounded
away from the axis, and Harnack's inequality with interior Schauder
estimates gives Eq.~\eqref{profile}. For a surface with unit outer
normal $\hat n$ and mean curvature $\hat H$ in the seed metric, the mean
curvature in $g=\Phi^4\hat g$ is
$\Phi^{-2}[\hat H+4\hat n(\ln\Phi)]$. In the limiting metric the torus
$\nu=\mathrm{const}$ has mean curvature
$-(d\,e^{-q}/a\,h_{k_c}^2)\,\mathcal T(\sigma)$, where, at $k=k_c$,
\begin{equation}
\begin{split}
 \mathcal T={}&\frac{3(1-k)^2(1-3\sigma^2)-2k(5-3k)\sigma^4}
        {3(1-k)\,\sigma R^3}\\
  &+\frac4{\sqrt k}\atanh\frac{\sqrt k\,\sigma}{R}
\end{split}
\end{equation}
is positive for $\sigma<\sigma_*=0.5938$, that is, $\nu<\nu_*=0.3710$.
For the sphere $r=\mathrm{const}$ the mean curvature has the sign of
$-\cos u\,[\cosh\nu+\sinh\nu\,\mathcal T]$, with $\cos u>0$ for $r>a$;
it is negative when $\nu\le\nu_*$ on the whole sphere, that is, for
$r\ge a\coth(\nu_*/2)=5.45\,a$.


\begin{thebibliography}{0}%
\makeatletter
\providecommand \@ifxundefined [1]{%
 \@ifx{#1\undefined}
}%
\providecommand \@ifnum [1]{%
 \ifnum #1\expandafter \@firstoftwo
 \else \expandafter \@secondoftwo
 \fi
}%
\providecommand \@ifx [1]{%
 \ifx #1\expandafter \@firstoftwo
 \else \expandafter \@secondoftwo
 \fi
}%
\providecommand \natexlab [1]{#1}%
\providecommand \enquote  [1]{``#1''}%
\providecommand \bibnamefont  [1]{#1}%
\providecommand \bibfnamefont [1]{#1}%
\providecommand \citenamefont [1]{#1}%
\providecommand \href@noop [0]{\@secondoftwo}%
\providecommand \href [0]{\begingroup \@sanitize@url \@href}%
\providecommand \@href[1]{\@@startlink{#1}\@@href}%
\providecommand \@@href[1]{\endgroup#1\@@endlink}%
\providecommand \@sanitize@url [0]{\catcode `\\12\catcode `\$12\catcode
  `\&12\catcode `\#12\catcode `\^12\catcode `\_12\catcode `\%12\relax}%
\providecommand \@@startlink[1]{}%
\providecommand \@@endlink[0]{}%
\providecommand \url  [0]{\begingroup\@sanitize@url \@url }%
\providecommand \@url [1]{\endgroup\@href {#1}{\urlprefix }}%
\providecommand \urlprefix  [0]{URL }%
\providecommand \Eprint [0]{\href }%
\providecommand \doibase [0]{https://doi.org/}%
\providecommand \selectlanguage [0]{\@gobble}%
\providecommand \bibinfo  [0]{\@secondoftwo}%
\providecommand \bibfield  [0]{\@secondoftwo}%
\providecommand \translation [1]{[#1]}%
\providecommand \BibitemOpen [0]{}%
\providecommand \bibitemStop [0]{}%
\providecommand \bibitemNoStop [0]{.\EOS\space}%
\providecommand \EOS [0]{\spacefactor3000\relax}%
\providecommand \BibitemShut  [1]{\csname bibitem#1\endcsname}%
\let\auto@bib@innerbib\@empty
\end{thebibliography}%


\begin{thebibliography}{99}
\bibitem{Brill} D. R. Brill, On the positive definite mass of the Bondi--Weber--Wheeler
 time-symmetric gravitational waves, Ann. Phys. (N.Y.) \textbf{7}, 466 (1959).
\bibitem{Wheeler} J. A. Wheeler, Geometrodynamics and the issue of the final state,
 in \emph{Relativity, Groups and Topology}, edited by C. DeWitt and B. DeWitt
 (Gordon and Breach, New York, 1964), pp. 317--522.
\bibitem{Eppley} K. Eppley, Evolution of time-symmetric gravitational waves:
 Initial data and apparent horizons, Phys. Rev. D \textbf{16}, 1609 (1977).
\bibitem{Holz} D. E. Holz, W. A. Miller, M. Wakano, and J. A. Wheeler, Coalescence of
 primal gravity waves to make cosmological mass without matter, in
 \emph{Directions in General Relativity, Vol.~2: Papers in Honor of Dieter
 Brill}, edited by B. L. Hu and T. A. Jacobson (Cambridge University Press,
 Cambridge, England, 1993), pp. 339--358.
\bibitem{KKS} J. Karkowski, P. Koc, and Z. \'Swierczy\'nski, Penrose inequality for
 gravitational waves, Classical Quantum Gravity \textbf{11}, 1535 (1994).
\bibitem{GarfinkleDuncan} D. Garfinkle and G. C. Duncan, Numerical evolution of
 Brill waves, Phys. Rev. D \textbf{63}, 044011 (2001).
\bibitem{Hilditch13} D. Hilditch, T. W. Baumgarte, A. Weyhausen, T. Dietrich,
 B. Br\"ugmann, P. J. Montero, and E. M\"uller, Collapse of nonlinear gravitational
 waves in moving-puncture coordinates, Phys. Rev. D \textbf{88}, 103009 (2013).
\bibitem{Hilditch17} D. Hilditch, A. Weyhausen, and B. Br\"ugmann, Evolutions of
 centered Brill waves with a pseudospectral method, Phys. Rev. D
 \textbf{96}, 104051 (2017).
\bibitem{BGW25} L. Bieri, D. Garfinkle, and J. Wheeler, Brill waves with slow
 fall-off towards spatial infinity, Classical Quantum Gravity \textbf{42},
 115011 (2025).
\bibitem{BGIMW26} L. Bieri, D. Garfinkle, J. Isenberg, D. Maxwell, and J. Wheeler,
 Asymptotically Euclidean solutions of the constraint equations with
 prescribed asymptotics, J. Math. Phys. (N.Y.) \textbf{67}, 102502 (2026).
\bibitem{AranhaOliveira} R. F. Aranha and H. P. de Oliveira, Similarity in the
 Arnowitt--Deser--Misner mass from Brill and Teukolsky initial datasets beyond
 the linear approximation, Phys. Rev. D \textbf{110}, 124058 (2024).
\bibitem{MastersonHobill} A. Masterson and D. Hobill, Critical parameterisation
 of the Brill wave initial value problem, \href{https://arxiv.org/abs/1809.05179}{arXiv:1809.05179}.
\bibitem{CantorBrill} M. Cantor and D. Brill, The Laplacian on asymptotically flat
 manifolds and the specification of scalar curvature, Compos. Math.
 \textbf{43}, 317 (1981).
\bibitem{BeigOM} R. Beig and N. \'O Murchadha, Trapped surfaces due to
 concentration of gravitational radiation, Phys. Rev. Lett. \textbf{66}, 2421 (1991).
\bibitem{BrillLindquist} D. R. Brill and R. W. Lindquist, Interaction energy in
 geometrostatics, Phys. Rev. \textbf{131}, 471 (1963).
\bibitem{Misner} C. W. Misner, The method of images in geometrostatics,
 Ann. Phys. (N.Y.) \textbf{24}, 102 (1963).
\bibitem{LinearBrill} I. Su\'arez Fern\'andez, T. W. Baumgarte, and D. Hilditch,
 Comparison of linear Brill and Teukolsky waves, Phys. Rev. D \textbf{104},
 124036 (2021).
\bibitem{companion} P. Koc, Analytic Brill wave initial data with trapped surfaces,
 companion paper, \href{https://arxiv.org/abs/2610.16773}{arXiv:2610.16773}.
\bibitem{Cohl} H. S. Cohl and J. E. Tohline, A compact cylindrical Green's function
 expansion for the solution of potential problems, Astrophys. J.
 \textbf{527}, 86 (1999).
\bibitem{BeigHusa} R. Beig and S. Husa, Initial data for general relativity with
 toroidal conformal symmetry, Phys. Rev. D \textbf{50}, R7116 (1994).
\bibitem{Husa98} S. Husa, Asymptotically flat initial data for gravitational wave
 spacetimes, conformal compactification and conformal symmetry,
 \href{https://arxiv.org/abs/gr-qc/9811005}{arXiv:gr-qc/9811005}.
\bibitem{Husa96} S. Husa, Initial data for general relativity containing a marginally
 outer trapped torus, Phys. Rev. D \textbf{54}, 7311 (1996).
\bibitem{Koc21} P. Koc, Analytical solution of Brill waves,
 \href{https://arxiv.org/abs/2107.02942}{arXiv:2107.02942}.
\bibitem{Natanzon} G. A. Natanzon, General properties of potentials for which
 the Schr\"odinger equation can be solved by means of hypergeometric
 functions, Theor. Math. Phys. \textbf{38}, 146 (1979).
\bibitem{AnderssonMetzger} L. Andersson and J. Metzger, The area of horizons and
 the trapped region, Commun. Math. Phys. \textbf{290}, 941 (2009).
\bibitem{Eichmair} M. Eichmair, The Plateau problem for marginally outer trapped
 surfaces, J. Differ. Geom. \textbf{83}, 551 (2009).
\bibitem{HuiskenIlmanen} G. Huisken and T. Ilmanen, The inverse mean curvature
 flow and the Riemannian Penrose inequality, J. Differ. Geom.
 \textbf{59}, 353 (2001).
\bibitem{KarkowskiTori} J. Karkowski, P. Mach, E. Malec, N. \'O Murchadha, and N. Xie,
 Toroidal trapped surfaces and isoperimetric inequalities, Phys. Rev. D
 \textbf{95}, 064037 (2017).
\bibitem{Bray} H. L. Bray, Proof of the Riemannian Penrose inequality using the
 positive mass theorem, J. Differ. Geom. \textbf{59}, 177 (2001).
\end{thebibliography}
\end{document}